\documentclass[prd,twocolumn,amsmath,amssymb,showpacs,floatfix,superscriptaddress,nofootinbib,preprintnumbers]{revtex4-1}

\usepackage{epsfig}
\usepackage{bm}
\usepackage{latexsym}
\usepackage{natbib}
\usepackage{url}
\usepackage{dcolumn}
\usepackage{color}
\usepackage[usenames,dvipsnames]{xcolor}
\usepackage{amsfonts,amssymb,amsmath}
\usepackage{graphicx,epsfig}
\usepackage{psfrag}
\usepackage{subfigure}
\usepackage{comment}
\usepackage[utf8]{inputenc}
\usepackage{amsmath}
\usepackage{hyperref}
\hypersetup{colorlinks=true}

\def\be{\begin{equation}}
\def\ee{\end{equation}}
\def\gev{{\rm \,Ge\kern-0.125em V}}

\begin{document}
 
\title{Warm Inflation in the light of Swampland Criteria}

\author{Suratna Das}%
\email{suratna@iitk.ac.in}
\affiliation{Department of Physics, Indian Institute of Technology, Kanpur 208016, India}

\date{\today}

\begin{abstract}
Warm Inflation seems to be the most befitting single-field slow-roll inflation scenario in the context of the recently proposed Swampland Criteria. We investigate the constraints these Swampland Criteria impose on Warm Inflation parameters and show that Warm Inflation is in accordance with both the current cosmological observations and the proposed Swampland Criteria in both weak and strong dissipative regimes depending on the value of the parameter $c$ which limits the slope of the inflaton potential according to the criteria. 
\end{abstract}

\maketitle

\section{Introduction}

At present, cosmic inflation \cite{PhysRevD.23.347, Starobinsky:1980te, Linde:1981mu, PhysRevLett.48.1220} is dominantly believed to be the mechanism which provides the initial seeds of inhomogeneties for the observed Cosmic Microwave Background anisotropies and the Large Scale Structures. According to the basic picture of cosmic inflation, this brief period of accelerated (quasi-)exponential (de Sitter) expansion of the Universe is driven by one slowly-rolling scalar field, the inflaton. Such exponential expansion leaves the Universe devoid of any matter at the end of inflation which calls for a period of reheating before the Universe re-enters the standard Big Bang (decelerating) phase. Based on when and how the Universe is `reheated', the basic inflationary mechanism bifurcates: the original picture where the inflaton field oscillates at the bottom of its potential at the end of inflation reheating the universe by dissipating its energy to a radiation bath (we will refer to this mechanism as `cold inflation'), and an alternative scenario where the inflaton field keeps dissipating its energy to a radiation bath during the course of inflation maintaining a constant radiation energy and thus avoids the conventional `reheating phase' at the end of inflation (this mechanism is known as Warm Inflation (WI) proposed in \cite{Berera:1995ie}).  

Inflation, believed to have taken place at the GUT scale or below, is assumed to be described by low-energy Effective Field Theories (EFTs). Such EFTs can be ultraviolet complete if they can be successfully embedded in a quantum theory of gravity, such as String Theory. String theory provides large landscapes where EFTs with Minkowski and anti-de Sitter vacua can be formulated with consistent quantum theory of gravity, whereas EFTs with de Sitter vacua lie in the surrounding `swamplands' where EFTs coupled to gravity render quantum theory of gravity inconsistent. This has led to a set of criteria, like the weak gravity conjecture \cite{ArkaniHamed:2006dz} and the recently proposed two Swampland Criteria \cite{Obied:2018sgi}, to ensure any (meta-)stable de Sitter vacuum EFT not to lie in the desired String landscapes. These two Swampland Criteria, barring de Sitter vacuum from String landscapes, pose potential threats to the basic mechanism of slow-roll inflation (as has been observed in \cite{Agrawal:2018own}) which we explain below. 

We state the two Swampland Criteria as proposed in \cite{Obied:2018sgi} and the reasons of concern raised by these criteria as far as inflationary dynamics is concerned:
\begin{itemize}
\item {\bf Swampland Criteria I ($\mathcal{SC}$I):} This criteria puts an upper bound on the field-range traversed by scalar fields in low-energy Effective Field Theories as 
\begin{eqnarray}
\frac{\Delta\phi}{M_{\rm Pl}}< \Delta,
\label{slc1}
\end{eqnarray}
where $\Delta\sim\mathcal{O}(1)$ and $M_{\rm Pl}$ is the reduced Planck mass.  This criterion emerges from the belief that there is a finite radius in field space of the EFT where the effective Lagrangian remains valid. At large distances $D$, generation of a tower of light scalar modes with masses 
\begin{eqnarray}
m\sim M_{\rm Pl}\exp(-\alpha D)
\end{eqnarray}
with $\alpha\sim\mathcal{O}(1)$, renders the validation of the effective Lagrangian \cite{Agrawal:2018own}. 

Lyth in his seminal paper \cite{Lyth:1996im} devised a lower bound on the range traversed by the inflaton field over the course of (single-field) inflation, dubbed the Lyth bound, which is related to the ratio ($r$) of amplitudes of the tensor and the scalar perturbations produced during inflation and is stated as \cite{Easther:2006qu}
 \begin{eqnarray}
\frac{\Delta\phi}{M_{\rm Pl}}\gtrsim\Delta N\sqrt{\frac r8},
\label{lythb}
\end{eqnarray}
where $\Delta N(\sim 60)$ is the number of e-folds of the duration of inflation. 
It is to note that to obtain sub-Planckian field-excursions during the course of inflation ($\Delta\phi\lesssim M_{\rm Pl}$), as demanded by the Swampland Criterion given in Eq.~(\ref{slc1}), it is required to have $r\lesssim\mathcal{O}(10^{-3})$ for $\Delta N\sim60$ \cite{Hotchkiss:2011gz}. The recent observations by PLANCK and BICEP2/KEK put an upper bound on tensor-to-scalar ratio as $r<0.064$ \cite{Akrami:2018odb}.

Generic polynomial scalar potentials, like quartic and quadratic which appear in Chaotic inflation models \cite{Linde:1981mu}, are known to be yielding way to large tensor-to-scalar ratios $(\mathcal{O}(10^{-1}))$ and hence are disfavoured by the current data \cite{Akrami:2018odb}. However, the `plateau models', like Higgs inflation \cite{Bezrukov:2007ep}, $R^2$ (Starobinsky) inflation \cite{Starobinsky:1980te}, pole inflation \cite{Broy:2015qna} and $\alpha-$attractor models \cite{Kallosh:2013hoa}, are known to be yielding such low tensor-to-scalar ratios ensuring small field excursions during the course of inflation and thus are not in much tension with $\mathcal{SC}$I.

\item {\bf Swampland Criteria II ($\mathcal{SC}$II):} The second Swampland Criterion puts a lower bound on the gradient of the scalar field potentials of any EFT as 
\begin{eqnarray}
M_{\rm Pl}\frac{|V'|}{V}\gtrsim c,
\end{eqnarray}
where $c\sim \mathcal{O}(1)$. Here prime denotes derivative w.r..t the inflaton field. It is shown in \cite{Obied:2018sgi} that the actual value of $c$ depends on the details of compactification and it often turns out to be of the order of $\sqrt{2}$ or greater in many string realizations and is not less than unity \cite{Kinney:2018nny}. However it has been argued in \cite{Kehagias:2018uem} that $c$ as small as $\mathcal{O}(10^{-1})$ doesn't go against perceiving de Sitter vacua in String landscapes. 

Single-field slow-roll inflationary dynamics, with canonical kinetic term and Bunch-Davies vacuum state, falls short in three different ways to meet this second Swampland Criteria:
\begin{enumerate}
\item First of all, the slow-rolling of the inflaton field is ensured by the flatness of its potential demanding the slow-roll parameters to be much smaller than unity. Thus the slow-roll condition, 
\begin{eqnarray}
\epsilon_\phi \equiv \frac{M_{\rm Pl}^2}{2}\left(\frac{V'}{V}\right)^2\ll1,
\end{eqnarray}
itself is in direct conflict with $\mathcal{SC}$II.
\item Secondly, single-field slow-roll inflation with canonical kinetic term and Bunch-Davies initial states predicts 
\begin{eqnarray}
r=16\epsilon_\phi.
\end{eqnarray}
Thus, $\mathcal{SC}$II demands that $r>8c^2$ which is in conflict with the 
current observational upper bound on $r$ as $r<0.064$ \cite{Akrami:2018odb} even if we consider $c$ to be as low as $10^{-1}$ \cite{Kehagias:2018uem}. This rules out the standard Chaotic inflation potentials\footnote{It has been recently argued in \cite{Lin:2018kjm} that Chaotic inflation on brane can be realised with polynomial potentials $V(\phi)\sim \phi^p$ for fine-tuned values of $p$ as $p\lesssim 0.35$} as well as the `plateau-like' potentials \cite{Bezrukov:2007ep, Starobinsky:1980te, Broy:2015qna, Kallosh:2013hoa} which were in accordance with the $\mathcal{SC}$I as discussed above. This problem can be avoided in cases of non-Bunch-Davies initial states \cite{Brahma:2018hrd, Ashoorioon:2018sqb} and non-canonical kinetic terms such as in $k-$inflation \cite{Das:2018hqy} models which, due to the modified dynamics, introduces a suppressing factor on the rhs of the above equation, alleviating the constraint on $r$. 
\item Lastly, we note that irrespective of the form of the potential during inflation, in the standard cold inflation scenario the inflaton field decays during reheating into the radiation bath while oscillating at the bottom of it potential, where $V'\sim0$. This is also in conflict with the second Swampland Criterion \cite{Agrawal:2018own}. 
\end{enumerate}
It has been recently noted in \cite{Das:2018hqy} that Warm inflation \cite{Berera:1995ie} suits $\mathcal{SC}$II the most. First of all, we note that as in WI the inflaton field dissipates to the radiation bath during the course of inflation, such dynamics brings in an additional friction term to the inflaton slow-roll equation of motion:
\begin{eqnarray}
3H\dot\phi+\Gamma\dot\phi\sim-V',
\end{eqnarray}
where $\Gamma$ is the decay rate of the inflaton field. This yields the slow-roll condition as 
\begin{eqnarray}
\epsilon_\phi\ll 1+Q,
\end{eqnarray}
where $Q=\Gamma/3H$. It is easily observed that in the strong dissipative regime ($Q>1$) $\mathcal{SC}$II can be easily met without hampering the slow-roll condition of the inflaton field. Secondly, we note that due to the modified inflaton dynamics WI predicts \cite{Berera:1999ws, Berera:2004vm, BasteroGil:2009ec, Bartrum:2013fia, Bastero-Gil:2016qru, Visinelli:2011jy, BasteroGil:2012zr, Bastero-Gil:2013nja, Visinelli:2014qla, Bastero-Gil:2018uep}
\begin{eqnarray}
r=\left(\frac HT\right)\frac{16\epsilon_\phi}{(1+Q)^{\frac52}},
\end{eqnarray}
where $T$ is the temperature of the radiation bath with $T>H$. Thus in the strong dissipative regime ($Q>1$) the suppressing factor $(H/T)/(1+Q)^{5/2}$ helps evade $\mathcal{SC}$II to remain in tune with the present observations. Lastly we note that as WI, by construction, doesn't call for a reheating phase at the end of inflation, it also doesn't call for potentials with $V'\sim0$ which further helps WI to remain in accordance with $\mathcal{SC}$II.
\end{itemize}
The aim of the present article is to further investigate WI scenario in the light of the Swampland Criteria to see how much these criteria constrain the parameters of WI, especially the parameter $Q$ which determines whether WI takes place in the weak $(Q<1)$ or strong $(Q>1)$ dissipative regime. 

However, before approaching to the main analysis of the paper, it is of importance to analyse the warm inflationary dynamics in some detail as the basic mechanism of WI differs from that of conventional `cold inflation' to some extent. The WI scenario demands coupling of the inflaton field with light degrees of freedom (DOF) to which the inflaton field would dissipate its energy to maintain a constant radiation bath throughout the course of inflation. The presence of such light DOFs can potentially modify the inflaton potential which drives inflation. As the Swampland Criteria are all about bounds on the form of the scalar (inflaton) potentials in any EFT, it is of importance of scrutinise how much the inflaton potential gets modified due to the presence of such light DOFs in a warm inflationary model. It was soon realized, after the proposal of warm inflation, that coupling the inflaton with such light DOFs during inflation is indeed a taxing job \cite{Yokoyama:1998ju}, and the reason for it is two-fold which can be easily understood if we consider Yukawa-like couplings $g\phi\bar\psi\psi$ of the inflaton field $\phi$ with light spinor fields $\psi$'s. First of all, the inflaton induces large masses to the sprinor fields $m_\psi=g\phi$, and in return the spinor fields induce large thermal corrections to the inflaton mass ($m_\phi\sim gT$) barring slow-roll for $T>H$. Both these hurdles can be overcome by fine-tuning the coupling $g$, which then goes against the effective dissipation of the inflaton field to the light DOFs and hence the scenario fails to sustain a constant thermal bath. This roadblock can be avoided in two circumstances. The first  way-out would be to allow the inflaton to couple to intermediate heavy scalars which will then eventually decay to the light DOFs. It was extensively estimated in \cite{Hall:2004zr} that in such warm inflationary scenarios the thermal correction to the inflaton potential is quite negligible within a global supersymmetry setup. WI with brane construction setup has also been studied in \cite{BasteroGil:2011mr} in this context and has been shown that the thermal corrections turn out to be Boltzmann suppressed. The other scenario where warm inflation can be successfully realised is the `Warm Little Inflaton' scenario \cite{Bastero-Gil:2016qru, Bastero-Gil:2018uep}, where the inflaton is treated as pseudo Nambu-Goldstone boson (corresponding to the relative phase between two complex Higgs scalars that collectively break a local $U(1)$ symmetry) coupled to a pair of fermionic fields through Yukawa interactions. The advantage of this scenario is that it bounds the masses of the fermions as $gM\cos(\phi/M)\bar\psi_1\psi_1$ and $gM\sin(\phi/M)\bar\psi_2\psi_2$ (with $M$ as the vacuum expectation value of the two Higgs scalar) and also, for $m_{\psi_{1,2}}\ll T$, the thermal mass correction due to the fermions cancel between the contributions of both the fermions, leaving only the subleading Coleman-Weinberg term. To achieve such cancellations discrete exchange symmetry is imposed in the scalar-spinor sector. The zero-temperature form of the potential is also protected agaianst large thermal correction due to the gauge symmetry of the underlying theory. Hence, in this case as well, the thermal corrections modify the slope of the inflaton potential negligibly and thus do not affect the slow-roll conditions of warm inflation scenario (see \cite{Bastero-Gil:2017wwl} for example) leaving the Swampland constraints intact for the warm inflationary case.

\section{Swampland Criteria and Warm Inflation}

We will first analyze $\mathcal{SC}$I to see what constraints it imposes on the WI parameter $Q$. As in cold inflation the $\mathcal{SC}$I is in contrast with the Lyth bound, as stated in Eq.~(\ref{lythb}), we need to determine the Lyth bound in the context of WI to appraise $\mathcal{SC}$I. We first note that the scalar power spectrum in WI receives two additive factors along with the form we get in cold inflation as \cite{Bartrum:2013fia}
\begin{eqnarray}
P_{\mathcal R}=\left(\frac{H}{\dot\phi}\right)^2\left(\frac{H}{2\pi }\right)^2\left[1+2n+\left(\frac TH\right)\frac{2\sqrt{3}\pi Q}{\sqrt{3+4\pi Q}}\right],\nonumber\\
\end{eqnarray}
where the last factor appears due to the presence of dissipative term in the inflaton equation of motion, and $n\equiv 1/(\exp(-k/aT)-1)$ denotes the Bose-Einstein distribution of the thermalized inflaton fluctuations. However, when $T>H$, a condition for thermal equilibrium, the scalar power spectrum can be approximated as \cite{BasteroGil:2009ec}
\begin{eqnarray}
P_{\mathcal R}\approx\frac{1}{8\pi^2\epsilon_\phi}\frac{H^2}{M_{\rm Pl}^2}(1+Q)^{5/2}\left(\frac{T}{H}\right).
\end{eqnarray}
The weakly coupled tensor modes, however, remain unaffected by the WI dissipative terms yielding the same tensor spectrum as in cold inflation :
\begin{eqnarray}
P_T=\frac{2H^2}{\pi^2M_{\rm Pl}^2}.
\end{eqnarray}
These yield the tensor-to-scalar ratio in the Warm Inflation scenario as 
\begin{eqnarray}
r=\left(\frac HT\right)\frac{16\epsilon_\phi}{(1+Q)^{\frac52}},
\label{r-eps-phi}
\end{eqnarray}
which can also be written as 
\begin{eqnarray}
r=\left(\frac HT\right)\frac{16\epsilon_H}{(1+Q)^{\frac32}},
\end{eqnarray}
as in WI
\begin{eqnarray}
\epsilon_H\equiv -\frac{\dot H}{H^2}\approx \frac{\epsilon_\phi}{1+Q}.
\end{eqnarray}
The modified inflaton dynamics of WI suggets
\begin{eqnarray}
\dot\phi=-\frac{\sqrt{2\epsilon_H}}{\sqrt{1+Q}}M_{\rm Pl}H,
\end{eqnarray}
which yields 
\begin{eqnarray}
\frac{\Delta\phi}{M_{\rm Pl}}=\sqrt{\frac r8\left(\frac TH\right)(1+Q)^{\frac12}}\Delta N,
\end{eqnarray}
rendering the modified Lyth bound for Warm inflation as \cite{Cai:2010wt, Visinelli:2016rhn}
\begin{eqnarray}
\frac{\Delta\phi}{M_{\rm Pl}}\gtrsim\sqrt{\frac r8\left(\frac TH\right)(1+Q)^{\frac12}}\Delta N.
\end{eqnarray}

$\mathcal{SC}$I suggests scalar field excursions (in Planck units) to be smaller than $\Delta\sim\mathcal{O}(1)$. But as the actual value of the parameter $\Delta$ is uncertain, we simply demand sub-Planckian field-excursions during the 60 e-folds duration of inflation. Thus demanding $\Delta\phi/M_{\rm Pl}<1$ we get
\begin{eqnarray}
r<\frac{8}{(\Delta N)^2 \sqrt{1+Q}}\left(\frac HT\right).
\end{eqnarray}
But the current observational bound suggests that $r<0.064$. These two conditions can be simultaneously met if 
\begin{eqnarray}
\frac{1}{(\Delta N)^2 \sqrt{1+Q}}\left(\frac HT\right)<0.008,
\end{eqnarray}
yielding 
\begin{eqnarray}
1+Q>\frac{1}{(0.008)^2(\Delta N)^4}\left(\frac HT\right)^2\sim 10^{-5},
\end{eqnarray} 
where we have considered $\Delta N \sim 60$ and $H/T\sim 10^{-1}$. We note that $\mathcal{SC}$I is easily satisfied by WI even in the weak dissipative regime $Q<1$. 
 
Let us now consider the second Swampland Criterion which calls for the following condition 
\begin{eqnarray}
\epsilon_\phi>\frac{c^2}{2}.
\end{eqnarray}
Combining this condition with Eq.~(\ref{r-eps-phi}) we get 
\begin{eqnarray}
r>\left(\frac HT\right)\frac{8c^2}{(1+Q)^{\frac52}}.
\end{eqnarray}
This condition satisfies the current observational upper bound on $r$ if 
\begin{eqnarray}
\left(\frac HT\right)\frac{c^2}{(1+Q)^{\frac52}}<0.008,
\end{eqnarray} 
yielding 
\begin{eqnarray}
1+Q>\left(\frac HT\right)^{\frac25}\left(\frac{c^2}{0.008}\right)^{\frac25}\sim 4,
\end{eqnarray}
where we have considered $c\sim \sqrt{2}$ and $H/T\sim 10^{-1}$. We note that $\mathcal{SC}$II puts a more stringent bound on $Q$ than $\mathcal{SC}$I. $\mathcal{SC}$II demands WI to take place in the strong dissipative regime $Q>1$. The upcoming observations, like COrE \cite{deBernardis:2017ofr} and LiteBIRD \cite{2018JLTP..tmp..124S}, will search for tensor-to-scalar ratio $r\sim\mathcal{O}(10^{-3})$ and a non-observance of $r$ by such missions would drive WI further deep into the strong dissipative regime. On the other hand, if $c$ turns out to be of the order of $10^{-1}$, as has been argued in \cite{Kehagias:2018uem}, then we can see that with the present bound on $r$ one gets $1+Q>0.5$. In such a situation, even weak dissipative regime WI scenarios would be in accordance with both the Swampland Criteria as well as with the current observations.

\section{Discussion and Conclusion} 

It was readily realised after the recently proposed Swampland Criteria \cite{Obied:2018sgi} that these criteria regarding formulation of UV-complete low-energy EFTs might have severe consequences for the cosmological epochs relying on de Sitter vacuum, like inflation and dark energy \cite{Agrawal:2018own}. Since then a series of papers has been written to counter such stumbling blocks and to put both inflation and dark energy back on track, a non-exhaustive list of such analysis would include \cite{ Achucarro:2018vey, Garg:2018reu, Dias:2018ngv, Kehagias:2018uem, Kinney:2018nny, Brahma:2018hrd, Das:2018hqy,  Lin:2018kjm, Ashoorioon:2018sqb}. The paper \cite{Das:2018hqy} enlists all the possible single-field slow-roll scenarios which are in accordance with the second Swampland Criterion and points out that Warm Inflation is the best option amongst all as far as one considers Swampland Criterion II. 

In this paper we analysed both the Swampland Criteria in the context of WI to investigate what constraints the criteria can impose on WI parameters, in particular $Q$ which serves as an indicator whether WI takes place in weak ($Q<1$) or strong ($Q>1$) dissipative regime. We find that $\mathcal{SC}$II puts more stringent bound on $Q$ than $\mathcal{SC}$I, implying that WI should take place in the strong dissipative regime with $Q>3$ when $c\geq \sqrt2$. However, if the parameter $c$, whose actual value depends on the methods of compactifications, can be brought down to as low as $10^{-1}$, then $\mathcal{SC}$II can also allow for weak dissipative regimes in WI as far as the present observations are concerned. We also note that if future observations lower the upper bound on tensor-to-scalar ratio then $\mathcal{SC}$II would push WI deeper into the strong dissipative regime. 

While this paper was under preparation, a similar analysis was presented in \cite{Motaharfar:2018zyb}. It is to note that the conclusion drawn in this paper differs from that of \cite{Motaharfar:2018zyb} as the later concludes that $Q$ should be of the order $\Delta N\sim 60$ or larger for WI to evade the Swampland Criteria, driving WI models very deep into the strong dissipative regime. This is of concern as it was shown in \cite{Motaharfar:2018zyb} that most of the models of WI do not allow for such large $Q$ as that would result in a redder scalar spectrum, which is not in accordance with the observation of the scalar spectral tilt $n_s$. Our conclusion differs from that of  \cite{Motaharfar:2018zyb} as we showed that the required value of $Q$ to be in tune with the Swampland Criteria is order of magnitude smaller than $\Delta N\sim 60$ and the Swampland Criteria can even allow for weak dissipative regime in WI if the parameter $c$ appearing in $\mathcal{SC}$II turns out to be smaller than unity (but positive). Hence, we note that WI still remains the best possibility amongst the single-field slow-roll inflation scenarios if the Swampland Criteria stand the test of time. 

\acknowledgements
The author would like to thank Arjun Berera for very useful discussions. 
 The work of S.D. is supported by Department of Science and Technology, 
 Government of India under the Grant Agreement number IFA13-PH-77 (INSPIRE Faculty Award).  
\label{Bibliography}
\bibliography{wi-sl}

\end{document}